\documentclass[aps,twocolumn,prd,dvipsnames,usenames,floatfix,superscriptaddress,nofootinbib,longbibliography]{revtex4-1}
\pdfoutput=1

\usepackage{aas_macros}

\usepackage[utf8]{inputenc}
\usepackage{graphicx}
\usepackage{amsmath}
\usepackage{amssymb}
\usepackage{bm}
\usepackage{paralist,array}
\usepackage{units} 
\graphicspath{{fig/}{./Plots/}} 
\usepackage{slashed}
\usepackage{todonotes}
\usepackage[colorlinks=true,citecolor=blue]{hyperref}

\graphicspath{{plots/}}

\newcommand{\eV}{\unit{eV}}

\newcommand{\GHz}{\unit{GHz}}

\newcommand{\K}{\unit{K}}
\newcommand{\ie}{\textit{i.e.}}
\newcommand{\eg}{\textit{e.g.}}
\newcommand{\zdec}{\ensuremath{z_{\rm dec}}}

\newcommand{\omgtwoone}{\ensuremath{\omega_{21}}}

\begin{document}

\title{Constraining dark photons and their connection to 21\,cm cosmology\\ with CMB data}

\author{Kyrylo Bondarenko}
\email{kyrylo.bondarenko@cern.ch}
\affiliation{Theoretical Physics Department, CERN, 1 Esplanade des Particules, Geneva 23, CH-1211, Switzerland}
\affiliation{L'Ecole polytechnique f\'ed\'erale de Lausanne, Route Cantonale, 1015 Lausanne, Switzerland}

\author{Josef Pradler}
\email{josef.pradler@oeaw.ac.at}
\affiliation{Institute of High Energy Physics, Austrian Academy of Sciences, Nikolsdorfergasse 18, 1050 Vienna, Austria}

\author{Anastasia Sokolenko}
\email{anastasia.sokolenko@oeaw.ac.at}
\affiliation{Institute of High Energy Physics, Austrian Academy of Sciences, Nikolsdorfergasse 18, 1050 Vienna, Austria}

\begin{abstract} 
     In the inhomogeneous Universe, the cosmological conversion of dark photons into ordinary photons (and vice versa)  may happen at a great number of resonance redshifts. This alters the CMB observed energy spectrum and degree of small-scale anisotropies. We utilize results from the \mbox{EAGLE} simulation to obtain the conversion probability along random line-of-sights to quantify these effects. We then apply our results to the case where dark photons are sourced by dark matter decay and their high-redshift conversion into ordinary photons modify the global 21~cm signal expected from the cosmic dawn era. Concretely, we show that a significant portion of the parameter space for which a converted population of photons in the Rayleigh-Jeans tail of the CMB explains the absorption strength observed by EDGES, is ruled out from the brightness temperature measurements of COBE/FIRAS and the CMB anisotropy measurements of Planck and SPT.
\end{abstract}
\maketitle 

\textbf{Introduction.}
The  observation of a global absorption feature in the sky, centered at $\nu_{\rm edg}=78$~MHz in frequency~\cite{Bowman:2018yin}, might mark the start of a new era in radio-astronomy. Albeit tentative in nature, if the result is correct, it will be the first observation of the cosmological era of first star formation (``cosmic dawn'').  The injection of \mbox{Ly-$\alpha$} photons by stars re-couples the spin-temperature of neutral hydrogen to the gas temperature, resulting  in a net absorption of 21~cm photons from the radiation bath~\cite{Furlanetto:2006jb}. The cosmological absorption feature at $\nu_0=1.4$~GHz frequency is then redshifted into the EDGES observational window centered at $\nu_{\rm edg}=\nu_0/(1+z_{\rm edg})$, where $z_{\rm edg}=17$. The amplitude of  absorption, measured in terms of a brightness temperature relative to  the radiation  background,
\begin{align}
\label{eq:prediction}
 T_{21}(\nu,z) \propto x_{\rm HI}(z) \left[1 - \frac{T_\gamma(z)}{T_s(z)} \right],
\end{align}
is reported as $T_{21}(\nu_{\rm edg},z_{\rm edg}) \simeq -0.5\K$~\cite{Bowman:2018yin}, twice the value expected from standard cosmology. In Eq.~(\ref{eq:prediction}), $x_{\rm HI}$ is the fraction of neutral hydrogen, $T_s$ is the spin temperature, and $T_\gamma$ is the temperature of the radiation background.

The result has triggered a fair amount of interest, as it may point to new physics operative at or right before the cosmic dawn. Most works have focused on lowering the value of $T_s$ by cooling the gas, \eg~through dark matter (DM) baryon interactions~\cite{Barkana:2018lgd,Barkana:2018qrx,Fialkov:2018xre,Fraser:2018acy,Liu:2019knx}, although these models are  often subject to severe constraints~\cite{Dvorkin:2013cea,Gluscevic:2017ywp,Xu:2018efh,Munoz:2018pzp,Berlin:2018sjs,DAmico:2018sxd,Barkana:2018cct}.  An alternative pathway may be a raised temperature of the radiation bath $T_\gamma$ in the vicinity of the 21~cm wavelength band at the redshift of formation~\cite{Feng:2018rje,Ewall-Wice:2018bzf,Pospelov:2018kdh}. In order to explain a twofold increase in absorption of 21~cm photons at $z_{\rm edg}$, the temperature of the radiation bath and hence the number-count of photons must be doubled at the relevant frequency.  Denoting by $\omega_{21}=2\pi \nu_0$, and by $\langle T_b \rangle$ the average brightness temperature of $\textit{extra}$ photons relative to the one from 
the CMB, 
\begin{equation}
\label{eq:explain-edges}
\langle T_b(\omgtwoone,z_{\text{edg}})\rangle \approx T_0 (1+z_{\rm edg})    
\end{equation}
must hold to explain the EDGES observed absorption amplitude. Here, $T_0$ is the temperature of the CMB today, $T_0=2.7260(13)\,\K$~\cite{Fixsen:2009ug}.
The relation above holds since the flux of CMB photons in the Rayleigh-Jeans (RJ) low-frequency tail at $z_{\rm edg}$, $\omega\ll T_0(1+z_{\rm edg})$, is given by $ dF_{\rm CMB}/d\omega d\Omega \approx (1+z_{\rm edg}) T_0 \omega/(4\pi^3)$ and temperature becomes a proxy for the number density and, by the same token, for the flux.

Reference~\cite{Pospelov:2018kdh} proposes that resonant conversion of dark photons into ordinary photons at $z_{\rm edg}\le z \le 1700$ leads to  the required non-thermal excess component of radiation in  the RJ tail of the CMB. For the resonance to occur, $m_{A'} = m_{A}(n_e)$ must hold, where $m_{A'}$ is the dark photon mass and $m_{A}(n_e)$ is the photon plasma frequency in a plasma with free electron number density $n_e$. The probability of conversion at resonance, $P_{A'\to A}(\omega) = 1 -  p$, is then regulated by the dark photon-photon kinetic mixing angle in vacuum, $\epsilon$, and a scale parameter $R$ that measures the degree of non-adiabaticity;  $\omega$ is the dark photon energy. In the present context, the resonances are non-adiabatic $p = \exp[- 2\pi \epsilon^2 m_{A'}^2 R/(2\omega)] \simeq 1$ and, hence, for a single resonance, $ P_{A'\to A}(\omega) \simeq \epsilon^2 m_{A'}^2 R/\omega $ with equal  probability for $A\to A'$ conversion.
Overall, this brings about a new connection between the physics of  sub-eV mass-scale dark states and  cosmology in the intermediate redshift interval between decoupling and reionization.%
\footnote{The cosmological and astrophysical viability of sub-eV dark photons has \eg\ been studied in \cite{Nelson:2011sf,Arias:2012az,Dubovsky:2015cca,Graham:2015rva,Kovetz:2018zes,Agrawal:2018vin,Wadekar:2019xnf,AlonsoAlvarez:2019cgw,McDermott:2019lch}}

The viability of the proposal hinges  on a pre-existing abundance and spectrum of dark photons prior to their conversion. One of the simplest possibilities is to source a population of $A'$ through DM decay, which for concreteness was chosen to be an axion-like particle with mass $m_a$ and decay mode $a\to A'A'$. The lifetime $\tau_a$ can be much larger than the age of the Universe and still yield a number of dark photons that is far in excess of CMB photons in the RJ tail, allowing for the possibility  $P_{A'\to A}(\omega) \ll 1$ and with it permissible small values on $\epsilon$.%
\footnote{Any appreciable strength of a direct photon decay mode $a\to AA$ is severely constrained by stellar energy loss arguments~\cite{Pospelov:2018kdh}.}
For $A'$ to be converted into  the 21~cm frequency band or above, the kinematic requirement on the mass reads $m_a \geq 2 \omega_{\rm 21}$, and in practice we investigate the range $m_a \sim 10^{-5} - 10^{-1}\text{ eV}$.  

\begin{figure}[t]
    \includegraphics[width=\columnwidth]{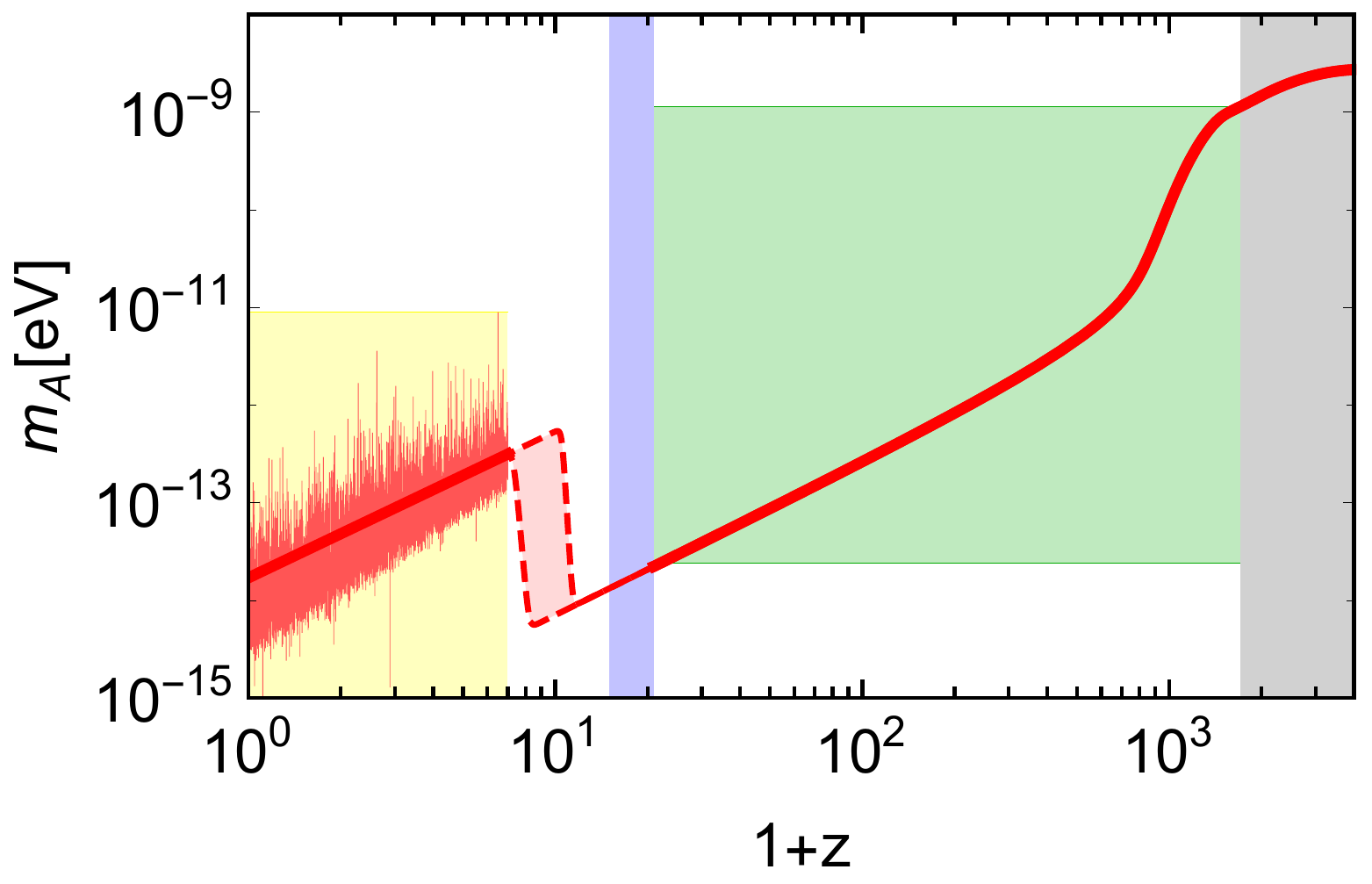}
    \caption{Plasma frequency as a function of redshift.
    For $z>20$, we use the average electron number density calculated from RECFAST~\cite{1999ApJ...523L...1S,Seager:1999km}, for $z<6$ we show the fluctuating electron number density  along a random line of sight extracted from the EAGLE simulation; the thick red line is the average density, and the dashed lines show the range in redshift associated with  hydrogen reionization~\cite{Heinrich:2016ojb,Aghanim:2018eyx}.  In the gray area, $z>1700$, converted photons are reabsorbed on matter. The blue shaded region shows the interval $20>z>15$, and the green shaded region the prospective range of $m_A$ relevant to EDGES. The intersected inhomogeneities have typically not broken away from the cosmological expansion as can be seen by their trend  with redshift.}
    \label{fig:resonanses}
\end{figure}

Further insight into the prospective parameter space comes from inspection of the resonance condition $m_A(n_e) = m_{A'}$. Up to negligible corrections from the neutral gas, the squared  plasma frequency is given by
\begin{equation}
        m_{A}^2(n_e) = \frac{4\pi \alpha n_e}{m_e} ,
\end{equation}
where $n_e$ is the free electron density, $\alpha$ is the fine-structure constant and $m_e$ is the electron mass. 
In the cosmological context and at high redshift, $n_e$ may to first approximation be treated as a homogeneous quantity that only depends of redshift, $n_e = n_e(z)$. Then, before reionization ($z_{\rm reion} \sim 10$) the plasma frequency is a monotonically falling function and resonant conversion before $z_{\rm edg}$ and after $z_{\rm abs} = 1700$---before which photons would be reabsorbed by inverse Bremsstrahlung~\cite{Chluba:2015hma}---may be achieved for $m_{A'} \sim 10^{-14} - 10^{-9}\,\eV$. As shown in Fig.~\ref{fig:resonanses}, during this period, the average plasma frequency scans through those values (green box) and the electron number density varies in the range  $n_{e, \text{res}} \in [3.5 \cdot 10^{-7}, 910]\text{ cm}^{-3}$.

It is the  purpose of this letter to point out that similar values of  electron density reoccur in the  late universe at $z\lesssim 6$, after reionization and when cosmological structures  have formed. This gives way to additional sources of creating extra photons relative to the CMB.  It is  illustrated in Fig.~\ref{fig:resonanses} where for $z<6$ we plot the plasma frequency based on the \textit{local} density of free electrons along a particular line-of-sight (LOS). The plot is made by extracting $n_e$ from the EAGLE simulation~\cite{Schaye2015,Crain2015}, the  procedure of which will be discussed in a companion paper~\cite{SimulationPaper}.  As can be seen, $m_A(z)$ becomes highly non-monotonic at low redshift and,
therefore, there may be many points along a typical LOS where the resonant condition is satisfied (even if the cosmological average value of $n_e$ is very different from the resonant value $n_{e,\text{res}}$).

\textbf{Conversion with inhomogeneities.} A late time converted photon flux may be detected against backgrounds in two different regimes: 
(i) in the range $n_{e, \text{res}} \in [10^{-9}, 10^{-1}]\text{ cm}^{-3}$
the resonant condition is satisfied many times in random directions. In this case, we deal with an all-sky signal, which,  however, is highly anisotropic as it correlates with large scale structure and accordingly has an imprint on the statistical properties of the CMB. 
(ii) For larger values of $n_{e,\text{res}}$, the resonant condition may be satisfied only in the centers of galaxies and galaxy clusters where the density of free electrons attains sufficiently large values that are not reached in the intergalactic medium. In this case, each such central region of resonant conversion would look like a localized source of the signal that can be observed by focused observations with high-resolution radio telescopes and be identified by their spectral properties. When choosing a random LOS on the sky, however, this second class of source is less likely to be met, and in the following we focus  our attention on possibility~(i).  

Consider a dark photon that was created from the decay of dark matter at redshift $z_{\text{dec}}$ with the energy $m_a/2$. It propagates towards Earth and at some redshift, $z_{\text{res}}$ resonantly converts into a normal photon.
The observed flux of such photons $F_{A}$  may be  written as
\begin{equation}
    \frac{d F_{A}}{ d\omega d\Omega}(\omega) = 
    \frac{d F_{A'}^{\text{no conv.}}}{d\omega d\Omega}(\omega) P^{\text{tot}}(\omega), \label{eq:flux_def}
\end{equation}
where $F_{A'}^{\text{no conv.}}$ is the flux of dark photons in the case of no conversion and $P^{\text{tot}}$ is the total probability that dark photons are converted into photons (and subsequently survive) while they propagate. 
In the limit that the DM lifetime is much larger than the age of the Universe, $\tau_a/t_0 \gg 1$, the relativistic dark photon flux $F_{A'}^{\text{no~conv.}}$ may be written as~\cite{Cui:2017ytb,Pospelov:2018kdh}
\begin{align}
    &\frac{d F_{A'}^{\text{no conv.}}}{d\omega d\Omega} (\omega) = 
    \frac{\Gamma_{a \to A' A'}}{2\pi H(z_{\text{dec}}) m_a} 
   \frac{\rho_{\text{DM}}(z_{\text{dec}})}{\omega(1+z_{\text{dec}})^3} \theta\left(\frac{m_a}{2} -\omega \right),  
    \label{eq:flux-dark-photon}
\end{align}
where $\rho_{\text{DM}}(z)$ is the DM density along the line of sight  and $\theta$ is the step function; we keep the dependence on the LOS direction implicit in our formulas. A central aspect of this work is that we take into account the inhomogeneous distribution of DM for $z<z_{\rm reion}$,  \ie\ $\rho_{\text{DM}} = \bar\rho_{\text{DM}} + \delta\rho_{\rm DM}$ where $\bar\rho_{\text{DM}} $ is the cosmological average DM density and $\delta\rho_{\text{DM}}  $ are the spatial flucutations around the mean. Finally, $H(z_{\rm dec})$ is the Hubble rate at decay. Note that in the 2-body decay there is a unique association  between energy and redshift of decay, $\zdec(\omega) = m_a/(2\omega) - 1$.

The total conversion probability $ P^{\text{tot}}(\omega)$ has two pieces, by assumption a guaranteed contribution at $z>z_{\rm edg}$ that modifies the cosmological 21 cm prediction, and a low redshift piece that amount to a (typically large) number of resonances at ``positions'' $z_i$
along the LOS for when $n_e(z_i)=n_{e,\text{res}}$ is met,
\begin{align}
       P^{\text{tot}}(\omega) =
      \frac{\pi \epsilon^2 m_{A'}^2}{\omega} & \left[ \frac{R_{\rm edg}}{1+z_{\rm res}} \theta(z_{\rm dec}(\omega)- z_{\rm res}) \right. \nonumber \\ & \left. + \sum_i \frac{R_i}{1+z_i} \theta(z_{\rm dec}(\omega)- z_i) \right].
    \label{eq:ptot}
\end{align}
It is the purpose of this paper to include the latter part in a way that is informed by cosmological N-body simulation and study its consequences. The scale parameter is given  by~\cite{Kuo:1989qe,Mirizzi:2009iz}
\begin{equation}
    R_i = \left|\frac{d \ln n_e}{d \ell}\right|^{-1}_{\ell = \ell(z_i)},
    \label{eq:Ri}
\end{equation}
where $\ell$ is a distance traveled by a photon along the line of sight.

An extra flux of photons at a particular frequency distorts the blackbody spectrum of CMB. Below, we will use the precision measurements by the COBE/FIRAS instrument~\cite{Fixsen:1996nj} to constrain these deviations.
In addition to an overall change in flux, fluctuations in the signal will be introduced by the variation of $n_e(z)$  and $\rho_{\rm DM}(z)$ along random LOS and according to their statistical properties over the sky.
This will imprint additional anisotropies of photons, and in particular in the CMB observational frequency window. 

As in both cases we are looking for the intersections of $n_e(z)$ with $n_{e,\text{res}}$, the details
of such LOS distributions are important. For this purpose, we use the results
obtained from numerical simulations in a companion paper~\cite{SimulationPaper}. There,  23 snapshots from $z=0$ to $6$ of comoving length 100~cMpc from the EAGLE simulation~\cite{Schaye2015,Crain2015} with a resolution up to $1\text{ ckpc}$ in the dense regions is used. Then $n_e(z)$ is extracted from those sample volumes with a resolution $0.25\text{ cMpc}$ and extrapolated in between those redshift anchor points.
Resonances at high redshift $z\geq z_{\rm edg}$ are expected to be  captured by the conversion probability computed from the single resonance given by $\langle m_A(z) \rangle = m_{A'}$~\cite{Pospelov:2018kdh} and which we follow here; the validity of this assumption will be further studied in~\cite{SimulationPaper}. In addition, the DM density distribution $\rho_{\text{DM}}(z)$ along random LOS was extracted from 37 snapshots from $z=0$ to $125$.

\textbf{EDGES normalization.}
In order to fix the normalization of the signal [proportional to the particle physics parameters $\epsilon^2 \Gamma_{a \to A' A'}$, see Eq.~\eqref{eq:flux_def}]
we impose (\ref{eq:explain-edges}), \ie\ we
require that the average number of converted photons with energy $\omgtwoone$ at redshift $z_{\text{edg}}$ is approximately equal to the number of CMB photons at that energy. Moving away from the RJ limit, it is beneficial to phrase the condition  (\ref{eq:explain-edges}) in terms of photon fluxes. One may write,
\begin{equation}
    \frac{dF_A}{d\omega d\Omega}(\omega) =
    \frac{\frac{dF_A}{d\omega d\Omega}(\omega)}{\left\langle\frac{dF_A}{d\omega d\Omega}(\omega_{21},z_{\text{edg}})\right\rangle}  \frac{dF_{\text{CMB}}}{d\omega d\Omega}(\omega_{21},z_{\text{edg}}).
    \label{eq:tmp_flux}
\end{equation}
The  ratio on the right hand side may be obtained by generalizing the equations \eqref{eq:flux_def}-\eqref{eq:Ri} to the conversion between $z_{\text{dec}}$ and $z_{\text{edg}}$ (rather than between $z_{\text{dec}}$ and  $z=0$). To this end note that the blue-shifted dark photon spectrum at redshift $z_{\text{edg}} $ reads,
\begin{equation}
    \frac{d F_{A'}^{\text{no conv.}}}{d\omega d\Omega} (\omgtwoone,z_{\text{edg}}) = (1 + z_{\text{edg}})^2\frac{d F_{A'}^{\text{no conv.}}}{d\omega d\Omega} \left(\omega'_{21} ,z=0 \right),
\end{equation}
where $\omega'_{21} = \omgtwoone/(1+z_{\rm  edg})$. From there we may rewrite the flux~\eqref{eq:tmp_flux} as,
\begin{align}
    \frac{d F_{A}}{d\omega d\Omega}(\omega) 
    = \frac{\omega T_0}{4 \pi^3} \times \mathcal{R}_H
    \times
   \mathcal{R}_{\text{DM}}
   \times
   \mathcal{R}_{\text{conv}},
    \label{eq:tbratio}
\end{align}
where
\begin{subequations}
\begin{align}
    \mathcal{R}_H &= \frac{\omega_{21}^{\prime\; 3}}{\omega^3}
    \frac{H[\zdec(\omega'_{21})]}{H[\zdec(\omega)]}  ,
    \label{eq:RH}
    \\
    \mathcal{R}_{\text{DM}} &=\frac{[1+\zdec(\omega'_{21})]^3 }{[1+\zdec(\omega)]^3}  \frac{ \rho_{\rm DM}[\zdec(\omega)]}{\langle \rho_{\rm DM}[\zdec(\omega'_{21})]\rangle}  \nonumber \\
    & = \frac{ \rho_{\rm DM}[\zdec(\omega)]}{\langle \rho_{\rm DM}[\zdec(\omega)]\rangle}
    ,
    \label{eq:RDM}
    \\
    \mathcal{R}_{\text{conv}} &=
    \left.  \left(\sum\limits_{z_i < z_{\text{dec}}(\omega)} \dfrac{R_i}{1+z_i}\right) \middle/\left(\dfrac{R_{\text{edg}}}{1+z_{\text{res}}}\right) \right. .
    \label{eq:Rconv}
\end{align}
\end{subequations}

\begin{figure*}[t]
    \centering
    \includegraphics[width=\columnwidth]{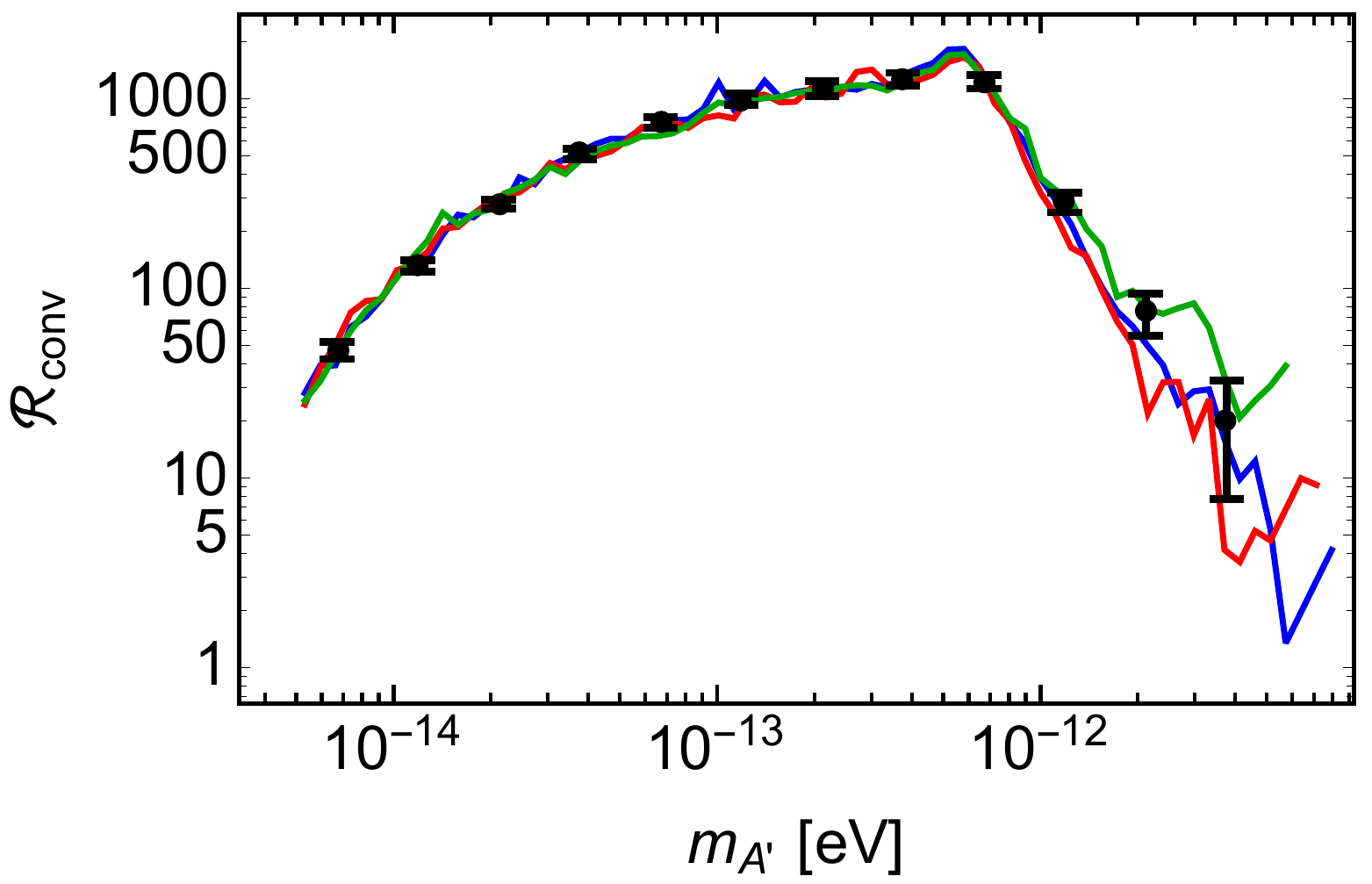}\hfill
    \includegraphics[width=0.96\columnwidth]{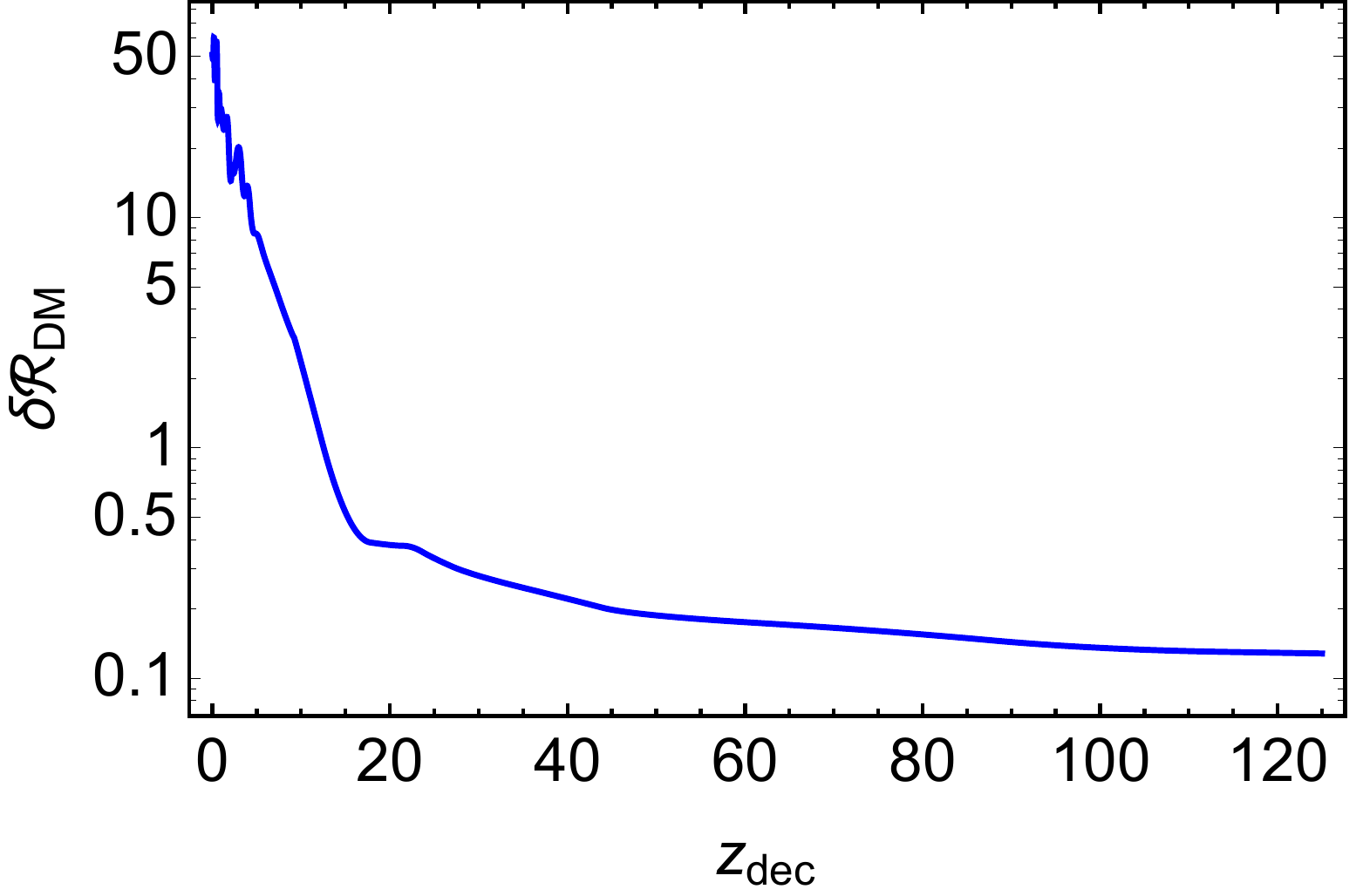}
    \caption{
    \textit{Left panel:} $\mathcal{R}_{\text{conv}}$ for 3 LOS (colored lines) as a function of  $m_{A'}$ for $z_{\text{dec}} = 20$.  The black dots with error bars are calculated using 100 random LOS and represent the fluctuations  $\delta \mathcal{R}_{\rm conv}$ that enter the calculation of the anisotropy in Eq.~(\ref{eq:tbratio2}).
    \textit{Right panel:}  Variation $\delta \mathcal{R}_{\text{DM}} $ of $\mathcal{R}_{\text{DM}}$ as a function of $z_{\text{dec}}$, calculated using 2900 random LOS from simulation; see~\cite{SimulationPaper}.}
    \label{fig:Rconv}
\end{figure*}

Above, $\mathcal{R}_H$ is a normalization factor fixed by the requirement that the model explains the EDGES-observed absorption strength, $\mathcal{R}_{\text{DM}}$ is the ratio of DM densities at redshift $z_{\text{dec}}(\omega)$ and the average DM density at the same redshift, and 
$\mathcal{R}_{\text{conv}}$ is the ratio of conversion probabilities with an overall energy dependence factored out. 
 The average of $\mathcal{R}_{\text{DM}}$ is equal to 1, as follows from its definition. Its  variation between different LOS can be sizable especially for small $z_{\text{dec}}$, as shown in the right panel of Fig.~\ref{fig:Rconv}  (see also~\cite{SimulationPaper} for details).
The quantity $\mathcal{R}_{\text{conv}}$ is a measure for the low-$z$ contributions that get added to the photon flux. At a given energy $\omega$, the primary dependence of $\mathcal{R}_{\text{conv}}$ is on $m_a$: the larger the DM mass is, the larger $\zdec$ becomes and the more resonances can be swept up. 
In the left panel of Fig.~\ref{fig:Rconv}  we show  $\mathcal{R}_{\text{conv}}$ as a function of the dark photon mass for 3 random LOS as well as its standard deviation (black error bars) obtained from 100 random LOS; for the latter, we have checked that the procedure converged.

\textbf{Blackbody distortions.}
 COBE/FIRAS~\cite{Fixsen:1996nj}  measured the spectral radiance $B_{\omega}(\omega)$ of the cosmological signal
in the frequency range 68 to 637~GHz with a precision of $10^{-4}$ and concordant with the Planck blackbody law. 
In the considered scenario there are two effects that modify the expected CMB spectrum. The first one is the conversion of dark photons that were created from DM decays. The second effect is a conversion of CMB photons to dark photons. Overall, we may write this as
\begin{align}
\label{Btot}
    B_\omega(\omega) = B^{\text{CMB}}_{\omega}(\omega) [1 -  P^{\text{tot}}(\omega)] + B^{A' \to A}_{\omega}(\omega),
\end{align}
where 
 $ B^{\text{CMB}}_{\omega}$ is the spectral radiance of the unmodified CMB, $P^{\text{tot}}(\omega)$ is given by~\eqref{eq:ptot} with the sum over all $z_i \leq 1700$, and $B^{A' \to A}_{\omega}(\omega)$ accounts for the extra photons created,
\begin{equation}
\label{eq:Bgain}
    B^{A' \to A}_{\omega}(\omega) = \frac{ \omega^2 T_0}{4\pi^3}
      \mathcal{R}_H
   \mathcal{R}_{\text{DM}}
   \mathcal{R}_{\text{conv}}.
\end{equation}
To compare these signals with the all-sky COBE/FIRAS data,  
we average over 100 random LOS in our simulation.

\begin{figure}[t]
    \centering
    \includegraphics[width=\columnwidth]{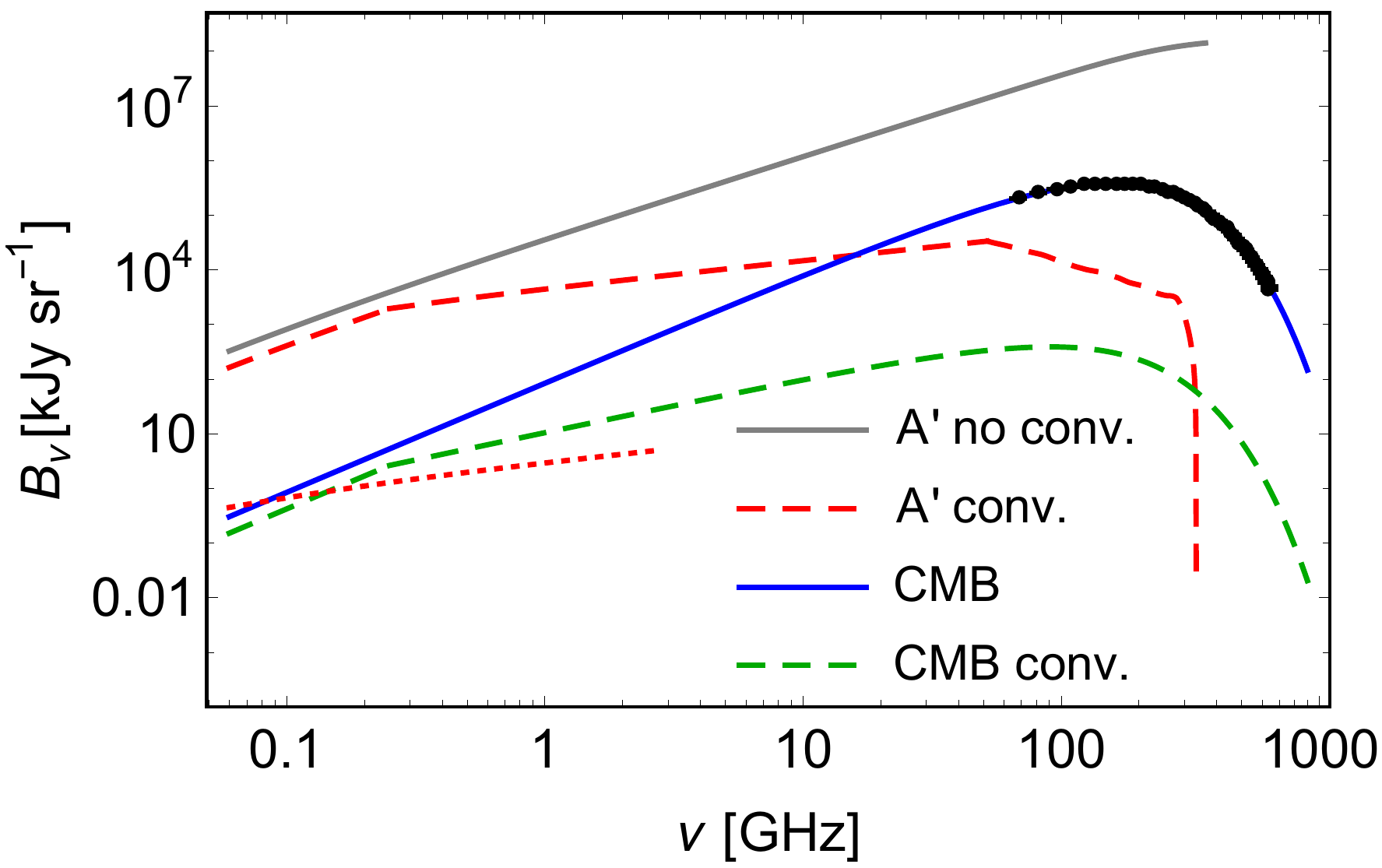}
    \caption{Spectral radiance of the unmodified CMB (blue line) and dark photons before conversion (gray line). The additional signal from converted dark photons, averaged over the sky is shown by the red dashed line, and converted CMB photons (lost into dark photons) are depicted by the green dashed line. The red dotted line shows the high-redshift contribution of the flux converted at $z>z_{\text{edg}}$, doubling the number of photons  of 21~cm wavelength at $z_{\rm edg}$.
    We fix DM and dark photon masses to $m_a=3\times 10^{-3}$~eV and $m_{A'}=3.8 \times 10^{-13}$~eV. Black points are data from COBE/FIRAS~\cite{Fixsen:1996nj}.}
    \label{fig:FIRAs_signal} 
\end{figure}

In Fig.~\ref{fig:FIRAs_signal} we show an exemplary signal for $m_a=3\times 10^{-3}$~eV and $m_{A'}=3.8 \times 10^{-13}$~eV.  
The spectral radiance of the unmodified CMB (dark photon flux) is shown by the solid blue (gray) line. The red dashed (dotted) line shows the  photon flux generated from dark photon conversion accounting for all resonances (the high redshift resonance at $z\geq z_{\rm edg}$). The dashed green line shows the dark photon contribution gained from converting CMB photons. 
As one can see, the number of photons lost through $A \to A' $ conversion is much smaller than the gain through  $A' \to  A $.
However, once the dark photon mass becomes large enough (above $10^{-11}$~eV), $A \to A' $ conversion also gives an important contribution, and, more generally, $A \to A' $ conversion itself can place an EDGES-independent bound on the presence of kinetically mixed dark photons~\cite{SimulationPaper}. In producing Fig.~\ref{fig:FIRAs_signal}, when the conversion probability reaches unity in (\ref{eq:ptot}), we take an arbitrary cut and set it to 0.5. This explains the kinks below 0.3~GHz. A more detailed treatment is inconsequential for the purpose of this work.

\begin{figure}[t]
    \centering
    \includegraphics[width=\columnwidth]{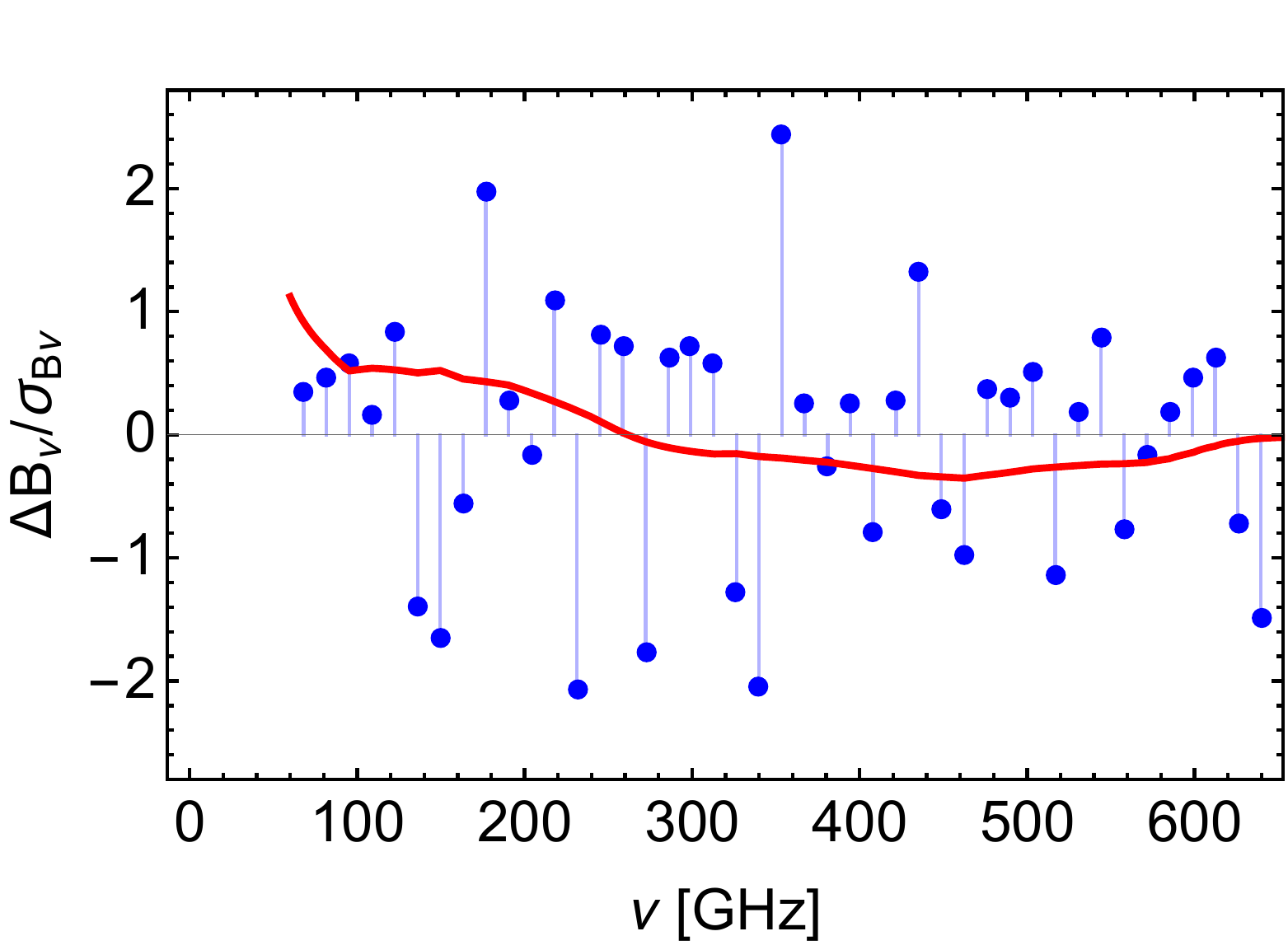}
    \caption{Residuals of COBE/FIRAS data $\Delta B_{\nu}$ against the best fit Planck spectrum normalized by experimental errors $\sigma_{B_{\nu}}$  (blue points) and an example of a $2\sigma$ excluded signal (red line) for $m_{a} = 3.8 \times 10^{-2}$~eV and $m_{A'} = 1.8\times 10^{-11}$~eV.}
    \label{fig:FIRAS_res}
\end{figure}

\begin{figure}[t]
    \centering
    \includegraphics[width=\columnwidth]{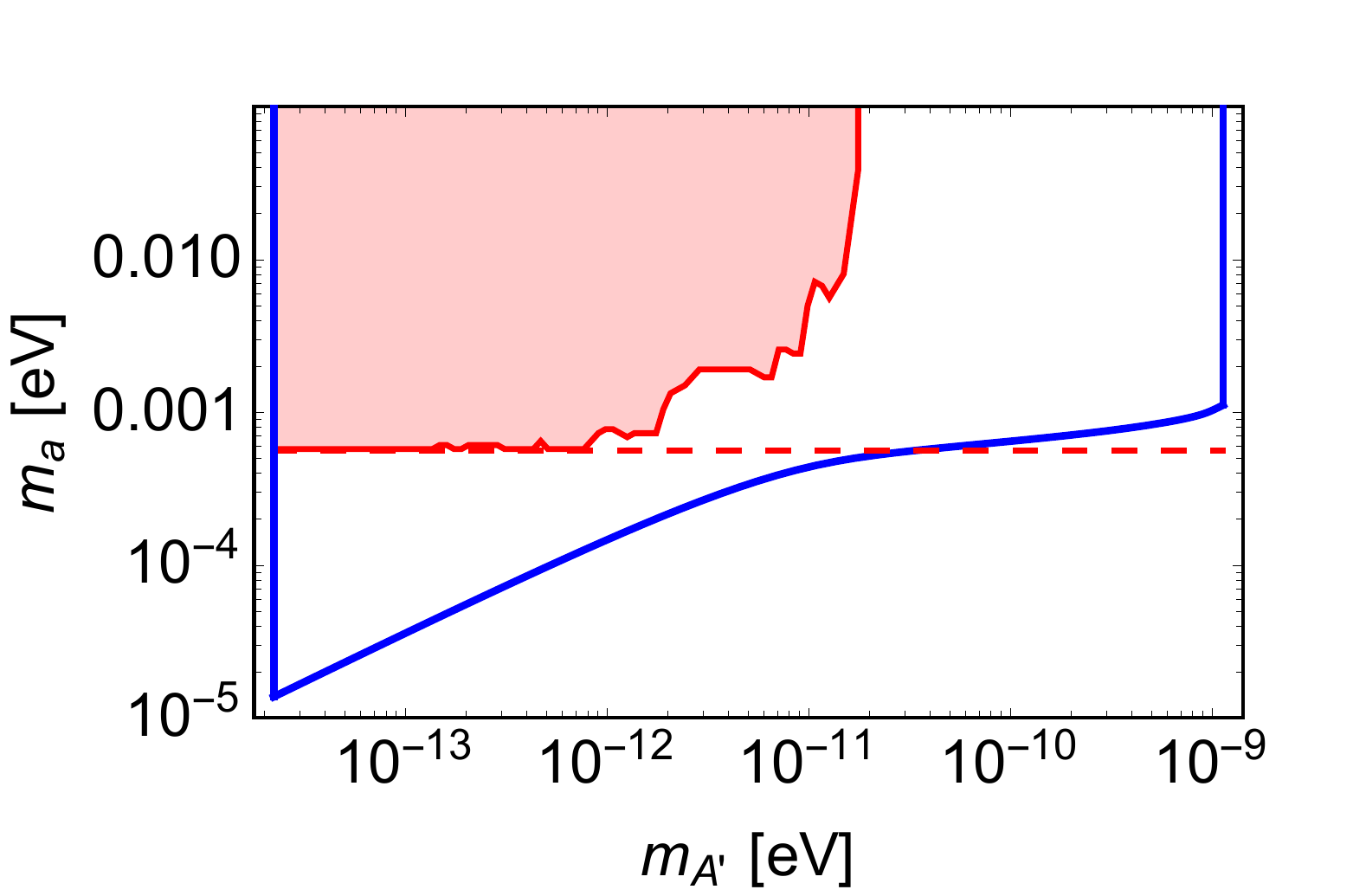}
    \caption{The red region is excluded at the $2\sigma$ level and above. from COBE/FIRAS data. The dashed red line shows the minimal DM mass that COBE/FIRAS can probe. In the region 
    inside the blue line, the EDGES anomaly can be explained.
    }
    \label{fig:FIRAS}
\end{figure}

In order to obtain a constraint in the $m_a$-$m_{A'}$ parameter space, we use (\ref{Btot}) to fit the data of COBE/FIRAS where we also allow the  temperature of the CMB, $T_0$, to float.  (We note though that its value remains within the standard errors of the original fit~\cite{Fixsen:1996nj}.)
Figure~\ref{fig:FIRAS_res} shows an example of an excluded parameter point against the residuals 
of COBE/FIRAS data; the fit of $B_\omega^{\rm CMB}$ has been subtracted for making the figure. 
Figure~\ref{fig:FIRAS} shows the excluded parameter region with  $2\sigma$ confidence level by the shaded area (red).
The dashed red line shows the minimum DM mass that can be probed by COBE/FIRAS. Below that line, injected photons fall below the longest wavelength bin. Conversely, the constraint vanishes for $m_{A'} \gtrsim 10^{-11}$~eV. Regions of higher electron densities are associated with small structures that are first typically not met through a random LOS approach and secondly not necessarily resolved because of averaging the LOS over a cross-sectional area of $0.25 {\rm Mpc}\times 0.025\,{\rm Mpc}$. A dedicated study of such ``point sources'' we leave for future work.

\textbf{Anisotropies.} 
As alluded to above, fluctuations in the signal flux are  introduced by the variations of $n_e$ and $\rho_{\rm DM}$ along the LOS.
In order to compute them, we now use the EAGLES-inferred $n_e$ and $\rho_{\rm DM}$ distributions to build a statistical sample by choosing random LOS through the sample volume. Whereas such  process masks any spatial correlations, it will still allow us to derive the overall variance in the expected signal. 
We find that the dispersion in the signal flux is significant and can be in excess of a few percent. 
This means that such a signal, provided that it is strong enough in absolute flux, can be excluded by any data on anisotropies, in particular from CMB measurements.

Like in the case of blackbody distortions of the CMB above, the constraining power in terms of mass of the progenitor will only be as good as the available frequency bands of the observed maps. For example,  the cosmological 21~cm signal is sensitive to radiation in the $\nu_0 = 1.4\,\GHz$ band requiring $m_a \geq 2\omega_{21} \simeq 12\mu\eV $, but the lowest frequency band of Planck is $\nu_{\text{low-band}}= 30\,\GHz$. Therefore, using Planck data, the scenario where dark photons explain EDGES can only be constrained for DM mass $m_a\geq 2\pi\nu_{\text{low-band}} = 250\,\mu\eV$.

Let us now estimate fluctuations in the flux induced by dark photon conversion. In Eq.~\eqref{eq:tbratio}, both quantities $\mathcal{R}_{\text{DM}}$ and $\mathcal{R}_{\text{conv}}$ depend on the LOS and can fluctuate,
\begin{align}
    \delta\left(\frac{d F_{A}}{d\omega d\Omega}(\omega)\right) 
    = \frac{\omega T_0}{4 \pi^3} \times \mathcal{R}_H
   \times
   \delta\left(\mathcal{R}_{\text{DM}} \mathcal{R}_{\text{conv}}\right),
    \label{eq:tbratio2}
\end{align}
where $\delta X^2 = \langle (X - \langle X\rangle)^2\rangle$
is the variance for different directions on the sky, along different lines of sight.  Note that the individual fluctuations $\delta \mathcal{R}_{\text{DM}}$ and $\delta\mathcal{R}_{\text{conv}}$ are largely uncorrelated when considered in the cosmological context: dark photons are sourced and converted at spatially well separated sites.%
\footnote{This is not true when considering the line-fluxes of individual systems, such as DM decay and dark photon conversion in a nearby cluster. Observational signatures of individual systems will be considered in a dedicated work.}
We therefore evaluate both quantities separately and use the formula for the variance of the product of two independent random variables $\mathcal{R}_{\text{DM}}$  and~$\mathcal{R}_{\text{conv}}$. 

The additional photons that are converted can lead to anisotropies that are in excess of the anisotropies as seen in the CMB. An overall measure of the latter is the
 all-sky variance of the CMB temperature, $\delta T_{\text{CMB}}^2$, which is obtained from the CMB temperature power spectrum,
\begin{equation}
\label{dTcmb}
    \delta T_{\text{CMB}}^2  = \sum_{\ell} \frac{2\ell + 1}{4\pi} C_{\ell}.
\end{equation}
For example,  using $C_{\ell}$ data from the Planck Legacy Archive~\cite{Akrami:2018vks}, we calculate the temperature fluctuations as,
\begin{equation}
    \delta T_{\text{CMB}} 
    \approx 1.1\times 10^{-4}\text{ K} \quad \Rightarrow \quad \frac{\delta T_{\text{CMB}}}{T_0} \approx 4\times 10^{-5}.
    \label{eq:deltaTCMB}
\end{equation}
Concordant values are obtained when using the power spectrum for the Planck $70$~GHz channel only~\cite{2018arXiv180706206P} and for SPT using the $150$~GHz band~\cite{article_SPT}.
Using temperature fluctuations we can calculate the variance in the CMB photon flux as
\begin{equation}
    \delta\left(\frac{d F_{\text{CMB}}}{d\omega d\Omega}\right) 
    = d\left(\frac{d F_{\text{CMB}}}{d\omega d\Omega}\right)/dT \times \delta T_{\text{CMB}},
\end{equation}
where we use the blackbody spectrum to calculate the derivative on the right-hand side.
 
As is well known, (\ref{dTcmb}) approximately implies that the contribution to $\delta T_{\rm CMB}^2$ per  $d\ln\ell$ is $D_l = \ell(\ell+1)/(2\pi)C_l$, \ie\ given by the angular power spectrum. 
When putting a constraint from CMB temperature anisotropies, one needs to take into account the angular resolution of the instrument.
For example, for Planck (SPT) at frequencies 70(150)~GHz, the highest reported multipole was $\ell_{\rm max} \approx 1250(11000)$ translating into an angular scale of $\theta_{\rm min}\approx \pi/\ell_{\rm max} \sim 10^{-3}(10^{-4})$. Whereas dark photon conversion will imprint anisotropies on a broad range of angular scales, unless the (cosmological) conversion happens nearby, the feature is expected to largely appear on small scales. For example, the Sunyaev-Zel’dovich (SZ) effect by hot intracluster gas is typically detectable as arcminute-scale CMB temperature fluctuations. In turn, dark photon conversion associated with SZ clusters may also imprint itself on similar angular scales $10^{-3}-10^{-4}$. It is important to note, however, that most of the intervening structure met along random LOS is not in collapsed structures, as it is less likely to intersect (clusters of) galaxies. This can  also be seen in the redshift scaling of the inhomogeneities shown in  Fig.~\ref{fig:resonanses} which  follow the $(1+z)^3$ trend; collapsed structures break away from the Hubble expansion.

The simplicity of using (\ref{dTcmb}) as a criterion to set a constraint is that one does not need to address the question of angular scale, as long as the simulation contributes features with  $\ell\lesssim \ell_{\rm max}$. Inherent in our approach of modeling the signal, however, is that the ``width'' of a random LOS may fall below the angular resolution of the instrument.
This means that more and more of our simulated LOS would contribute to a single ``angular pixel'' of the instrument. In other words, an averaging procedure is necessary where the number of LOS is such that a pixel is filled, washing out part of the anisotropy; in addition, this approach is computationally more expensive.
Here, to take this into account we estimate the LOS number $N_{\rm LOS}(z_{\text{dec}})$  with a  cross section $25\text{ ckpc}\times 250\text{ ckpc}$ per LOS~\cite{SimulationPaper} that is needed to fill  a pixel of Planck (SPT). We assume that fluxes given by each LOS are independent and divide the standard deviation of the flux calculated by Eq.~\eqref{eq:tbratio2} by a factor $\sqrt{N_{\rm LOS}(z_{\text{dec}})}$. This is expected to approximate the full simulation well as we confirm a scaling  of the anisotropy with  $1/\sqrt{N_{\rm LOS}(z_{\text{dec}})}$ in our numerical simulation.

\begin{figure}[t]
    \centering
    \includegraphics[width=\columnwidth]{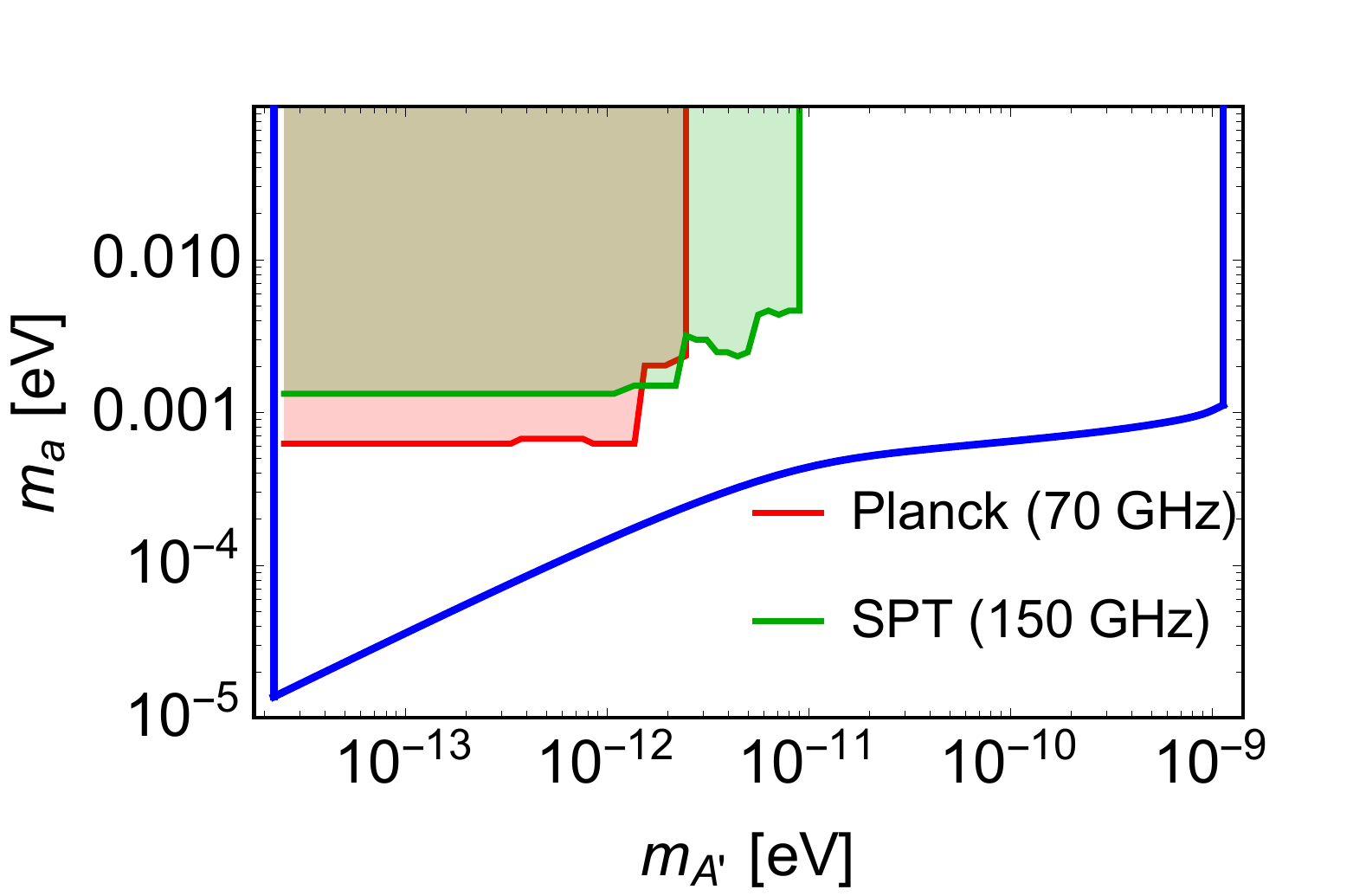}
    \caption{
    Shaded regions depict the area in the $m_a-m_{A'}$ parameter space where the temperature anisotropy induced by $A’\to A$ conversion is expected to be larger than (\ref{dTcmb}) at one of the Planck frequencies at $70$~GHz (red color) and one of the SPT frequencies at $150$~GHz (green color). We estimate the anisotropy from (\ref{eq:tbratio2}) with the variance extracted from 100 random LOS. As the LOS number $N_{\text{LOS}}(z)$ that can contribute to one ``angular pixel'' of the instrument (with a minimum angle of $ 10^{-3}$ for Planck and $10^{-4}$ for SPT) can be larger, at every redshift we penalise the contribution to the signal anisotropy by dividing by $\sqrt{N_{\text{LOS}}(z)}$. This estimate demonstrates that Planck and SPT can exclude a significant part of the parameter space when the proposal of~\cite{Pospelov:2018kdh} is used to explain the EDGES anomaly (inside the blue line).
    }
    \label{fig:Planck_res}
\end{figure}

The result of this procedure is shown in Fig.~\ref{fig:Planck_res}. Using the all-sky variance $\delta T/T_0$ from \eqref{eq:deltaTCMB} as a criterion, shows that Planck and SPT are expected to exclude a significant part of the parameter space where parameters are fixed to the EDGES absorption strength (region inside the blue line). For the constraint we evaluate (\ref{eq:tbratio2}) by picking two exemplary frequencies, 70~GHz (Planck) and  150~GHz (SPT) at which the CMB is well observed.
For small values of DM mass, corresponding to low $\zdec(70~\GHz)$ or $\zdec(150~\GHz)$, the exclusion is driven by DM density fluctuations, $\delta \mathcal R_{\rm DM}$, while for large DM mass the anisotropy induced by $\delta \mathcal R_{\rm conv}$ becomes important.
We expect that a more elaborate treatment that takes into account correlations between near LOS together with an angle-resolved version of~\eqref{eq:deltaTCMB} [with summation restricted over a range of $\ell$-modes] will yield a greater strength of exclusion. A detailed study of this will be presented elsewhere.

\textbf{Conclusions and outlook.} In this work we take the first step towards a realistic modeling of dark photon conversion in the low-redshift inhomogeneous Universe. Using the EAGLE simulation we extract the electron number density along random LOS, building a representative sample of in-flight probabilities for resonant photon-dark photon conversion (in both directions). We demonstrate the effect and importance of late time dark photon conversion by studying ensuing constraints on the proposed model \cite{Pospelov:2018kdh} for explaining the EDGES observed absorption strength of 21~cm radiation at redshift $z_{\rm edg}$. Concretely, we fix the product of squared kinetic mixing parameter and the decay rate of dark matter into dark photons such, that a high-redshift resonance at $z>z_{\rm egd}$ yields an extra photon population at the hydrogen hyperfine transition  energy at $z_{\rm edg}$  equal to the number of CMB photons at that redshift. This resonance condition requires the dark photon mass to be in the range $10^{-14}\lesssim m_{A'}/\eV\lesssim 10^{-9}$. We then find that the prospective region $m_{A'} \lesssim 10^{-11}\,\eV $ is largely excluded by late-time $A'\to A$ conversion after reionization once the progenitor mass is high enough to spill photons into the first COBE/FIRAS bin,  $m_a \gtrsim 6\times 10^{-4}\,\eV$. We find that the  enhancement in the cosmological flux relative to the high-redshift resonance is typically significant, and can be up to three orders of magnitude. 

In a second part we then advocate for using anisotropy measurements as a diagnostic tool to constrain late-time dark photon conversion. Additional temperature fluctuations are imprinted through variations of electron and dark matter densities along the line of sight. These two contributions are \textit{a priori} uncorrelated when decay- and conversion-redshift differ. We find that either contribution can imply a  variation of $\delta T/ T_0 $ in  excess of observed CMB anisotropies.
In the concretely studied setup, the region of parameter space that is disfavored through anisotropies is similar to the one excluded by the absolute flux measurements as discussed above. However, studying anisotropies in the signal makes the method amenable to be applied to a larger set of observations as an absolute calibration of the instrument is not required.

There are several directions for improvement and further investigations:
\begin{itemize}
\setlength\itemsep{0cm}
\item[-] More precise constraints from CMB anisotropies can be derived by taking a LOS approach that includes correlations that respect the finite angular resolution of the instruments. To this end, note that a sample volume of 100~cMpc should in principle allow for this as its minimum angular size on the sky is 0.01 for the studied redshift range, well above the resolution of Planck and SPT.  In addition, one can better constrain features imprinted on certain angular scales by using information of the power spectrum that is reflective of the angular size (rather than using the all-sky variance, as we conservatively did in this work.)

\item[-]  The highest electron densities that are resolved through our averaging procedure along LOS over a transverse area of $0.25~{\rm cMpc}\times 0.025~{\rm cMpc}$ are in the ballpark of $10^{-1}\,{\rm cm^{-3}}$,
that is close the densities associated with the centers of clusters of galaxies (although we typically do not intersect them.)
Hence, to test the region $m_{A'}\gtrsim 10^{-11}\,\eV$, we need to concentrate on dense, compact systems such as individual galaxies and our (random) LOS should be replaced by fine-grained versions with directions that intersect their central regions. In turn, dedicated observations of those regions rather than all-sky data like Planck (and the corresponding high-resolution simulations) can be used. 

\item[-] If the signal is present at $70$~GHz, it will also be present at lower frequencies, \eg\ in the Planck LFI instruments' 30~GHz and 44~GHz bands. Moreover, according to Eq.~\eqref{eq:tbratio2}
the overall amplitude of anisotropy will be enhanced due to the factor $\mathcal{R}_H$ which increases with decreasing frequency.
A cross-correlation of signals will help to distinguish it from foregrounds that are prevalent in low-frequency maps. Moreover, the observations are not constrained to be showing the CMB. Lower global frequency maps---albeit foreground dominated---exist and can be exploited in this context. In addition, the measurements of extra-Galactic sky temperature at frequencies below COBE/FIRAS can also be used to constrain excess flux contributions, most notably by ARCADE-II starting from 3~GHz~\cite{2011ApJ...734....5F}, or the Haslam map at 480~MHz~\cite{1981A&A...100..209H}.

\item[-] 
In this paper we restrict the study of inhomogeneous dark photon conversion to $z\leq 6$. Thereby we avoid the epoch of hydrogen reionization whose details remain uncertain. It is clear, however, that the redshift interval centered around $z_{\rm reion}$ can be a substantial source of additional anisotropies due to the patchy nature of the reionization process (see \eg~\cite{2017MNRAS.466..960P}), warranting a dedicated study. 

\item[-] Finally, it is important to note that even in the absence of a pre-existing dark photon flux, photons may convert to  dark states whenever the resonance condition is met. This will likewise imprint additional anisotropies onto the CMB and generally lead to flux modifications, implying constraints in the $\epsilon$-$m_{A'}$ parameter space that are independent of the dark matter embedding. Hence our study has a broader underpinning, as the LOS photon ``optical depth'' against conversion can be important in more general contexts.

\end{itemize}

These issues will in part be addressed in an upcoming companion paper~\cite{SimulationPaper} as well as in future work. The study of dark photons and their conversion in the cosmological context has by now a long history, but the role of inhomogeneities is only starting to be addressed. This work takes a first step showing that because the resonance condition can be simultaneously met at high- and multiple low-redshift locations, the latter must be included in any phenomenological study of cosmological dark photon  propagation.

\paragraph*{Acknowledgements.} We thank  A.~Arámburo García, A.~Boyarsky,  S.~Ploeckinger for collaboration
and S.~Sarkar for useful discussions. This research was supported in part by the National Science Foundation under Grant No.~NSF PHY-1748958. AS and JP are supported by the New Frontiers program of the Austrian Academy of Sciences. KB is supported by the European Research Council (ERC) Advanced Grant ``NuBSM'' (694896).

\textit{Note added}: When this paper was at the final stage of its preparation,~\cite{Caputo:2020bdy} appeared where a similar approach to describe late time photon-dark photon conversion was developed independently. The main difference between the two methods is that in the current paper we use simulated LOS, while in~\cite{Caputo:2020bdy} an average probability distribution function (also extracted from numerical simulations) is used.

\bibliographystyle{JHEP}
\bibliography{ship.bib}

\end{document}